\documentclass[aps,prl,twocolumn,superscriptaddress,showpacs]{revtex4}
\pdfoutput=1
\usepackage{graphicx}
\usepackage{mathrsfs}
\usepackage{bm}
\usepackage{amsmath}
\usepackage{dcolumn}
\usepackage{epstopdf}
\begin{document}
\title{Valley-Polarized Quantum Anomalous-Hall Effects in Silicene}

\author{Hui Pan}
\affiliation{Department of Physics, Beihang University, Beijing 100191, China}
\author{Zhenshan Li}
\affiliation{Department of Physics, Beihang University, Beijing 100191, China}
\author{Cheng-Cheng Liu}
\affiliation{School of Physics, Beijing Institute of Technology, Beijing 100081, China}
\affiliation{Beijing National Laboratory for Condensed Matter Physics,
 Institute of Physics, Chinese Academy of Sciences, Beijing 100190, China}
\author{Guobao Zhu}
\affiliation{Beijing National Laboratory for Condensed Matter Physics, Institute of Physics, Chinese Academy of Sciences, Beijing 100190, China}
\author{Zhenhua Qiao}
\affiliation{Department of Physics, The University of Texas at Austin, Austin, Texas 78712, USA}\affiliation{ICQD, Hefei National Laboratory for Physical Sciences at Microscale, University of Science and Technology of China, Hefei, Anhui 230026, China}
\author{Yugui Yao}
\email{ygyao@bit.edu.cn}
\affiliation{School of Physics, Beijing Institute of Technology, Beijing 100081, China}

\begin{abstract}
We demonstrate a new quantum state of matter---valley-polarized quantum anomalous Hall effect in monolayer silicene. In the presence of extrinsic Rashba spin-orbit coupling and exchange field, monolayer silicene can host a quantum anomalous-Hall state with Chern number $\mathcal{C}=2$. We find that through tuning Rashba spin-orbit coupling, a topological phase transition gives rise to a valley polarized quantum anomalous Hall state, i.e. a quantum state which exhibits electronic properties of both quantum valley-Hall effect (valley Chern number $\mathcal{C}_V=3$) and quantum anomalous Hall effect with $\mathcal{C}=-1$.
\end{abstract}


\maketitle
Silicene, the counterpart of graphene~\cite{Novoselov} for silicon, has a honeycomb geometry and low-buckled structure. Due to its special Dirac-cone structure in low-energy spectrum and promising applications in nanoelectronics~\cite{Lalmi,Vogt,Fleurence,LanChen,W.F.Tsai,Liu1}, silicene has attracted much attention both theoretically and experimentally. Distinct from graphene, silicene possesses a stronger intrinsic spin-orbit coupling and a considerable bulk band gap can open at the Dirac points. Therefore, silicene becomes a good candidate to realize the quantum spin Hall state~\cite{Liu1,Liu2}, a quantized response of a transverse spin current to an electric field~\cite{Kane1,Kane2,Hasan,X.-L.Qi}. On the other hand, a tunable extrinsic Rashba spin-orbit coupling from the mirror symmetry breaking about the silicene plane destroys this effect~\cite{Kane1,Kane2}. Interestingly, this helps establish another striking topological phase --- quantum anomalous Hall effect (QAHE)~\cite{Nagaosa,Tse,Onoda,JunDing,Qiao1,RuiYu,Chang}. Unlike the quantum Hall effects from Landau-level quantization, QAHE originates from the joint effects of spin-orbit coupling and local magnetization. Although QAHE has been proposed for over twenty years, there was only one evidence of realization in a magnetic topological insulator until recently~\cite{Chang}.

Similar to real spin, valleys $K$ and $K'$ in honeycomb structures provide another tunable binary degrees of freedom to design valleytronics. By breaking the inversion symmetry, e.g. the introduction of staggered AB sublattice potentials in honeycomb lattices, a bulk band gap opens to host a quantum valley-Hall effect (QVHE) \cite{Lu,Qiao3,Yao1,Yao2,Xiao2} characterized by a valley Chern number $\mathcal{C}_V=\mathcal{C}_K-\mathcal{C}_{K'}$. Since both intrinsic and extrinsic spin-orbit couplings in silicene are comparable, one can expect that these effects should result in new topological phenomena. It has turned out that manipulating the spin-orbit couplings is one of the most effective means to control the properties of silicene. For example, an interesting valley-polarized metallic state has been reported in silicene recently~\cite{Ezawa1}. Because the intrinsic and Rashba spin-orbit couplings give different responses on the conduction and valence bands of silicene, it is natural to imagine that a topological phase including both properties of QAHE and QVHE may exist in silicene.

In this Letter, we report a valley-polarized quantum anomalous Hall phase in monolayer silicene. When the time reversal symmetry is broken from the exchange field, we find that the combination of intrinsic and Rashba spin-orbit couplings can result in a new topological phase, i.e., a bulk band gap closing and reopening occurs along with the increasing of Rashba spin-orbit coupling at fixed intrinsic spin-orbit coupling and exchange field. Through analyzing the resulting Berry curvatures for the occupied valence bands, we show that the nonzero Chern number indicates a quantum anomalous-Hall phase. Surprisingly, we further find that valleys $K$ and $K'$ give rise to different Chern numbers, i.e. $\mathcal{C}_K=1$ and $\mathcal{C}_{K'}=-2$. This imbalance signals a quantum valley-Hall phase with valley Chern number $\mathcal{C}_V=3$. Our finding not only provides a platform to design low-power electronics but also advances the application of silicene-based valleytronics.

\begin{figure*}
\includegraphics[width=17cm]{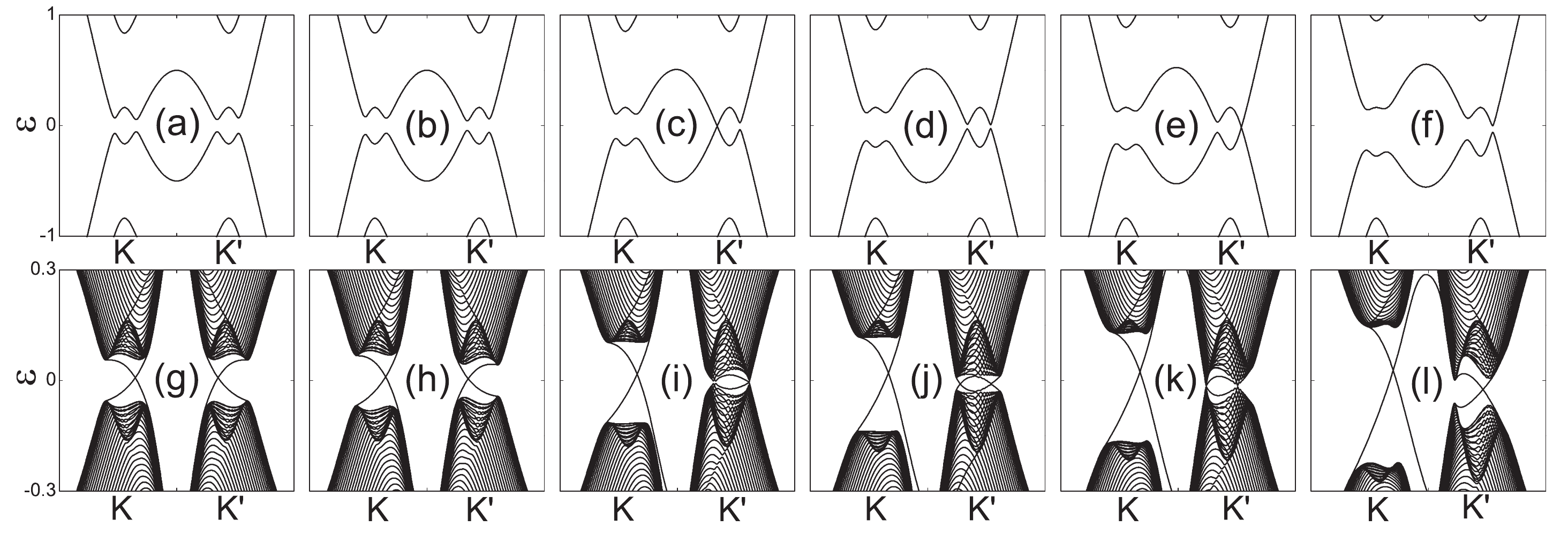}
\caption{\label{FIG:Ek-bulk}(Color online) Evolution of band structures of the bulk [panel (a)-panel (f)] and zigzag-terminated [panel (g)-panel (l)] silicene as a function of Rashba spin-orbit coupling $t_R$ at fixed intrinsic spin-orbit coupling $t_{SO}$ and exchange field $M$. (a) $t_R=0$. Bulk energy gaps open around the $K$ and $K'$ Dirac points. The size of bulk gap near valley $K$ is exactly the same as that near valley $K'$. Panels (b)-(f): $t_R=0.01,0.044,0.06,0.08,0.12$, respectively. Along with the increasing of $t_R$, the bulk gap around valley $K$ gradually increases, while the bulk gap near valley $K'$ closes twice [see panels (c) and (e)] and reopens twice [see panels (d) and (f)]. In panels (g)-(l), the valley-associated gapless edge modes at valley $K$ is unchanged, but those for valley $K'$ changes, i.e. there are two/one pairs edge modes after the bulk gap reopens. Other parameters are set to be $t_0/t=0.07$, $t_{SO}/t=0.10$, and $M/t=0.50$.}
\end{figure*}

In the tight-binding approximation, the Hamiltonian for silicene in the presence of spin-orbit couplings and exchange field can be written as~\cite{Liu1,Liu2}:
\begin{eqnarray}
\label{EQ:TBH}
H&=&-t\sum_{\langle ij \rangle \alpha}c^\dagger_{i\alpha}c_{j\alpha}
 +i t_{0}\sum_{\langle\langle ij  \rangle\rangle \alpha\beta}\nu_{ij}c^\dagger_{i\alpha}{\bm \sigma}^{z}_{\alpha\beta}c_{j\beta} \nonumber \\
 &-&i t_{SO}\sum_{\langle\langle ij \rangle\rangle \alpha\beta}\mu_{ij}c^\dagger_{i\alpha}
  (\vec{\bm \sigma}\times \hat{d}_{ij})^{z}_{\alpha\beta} c_{j\beta} \nonumber\\
 &+&i t_{R}\sum_{\langle ij \rangle \alpha\beta}c^\dagger_{i\alpha}
  (\vec{\bm \sigma}\times \hat{d}_{ij})^{z}_{\alpha\beta} c_{j\beta}
 +\emph{M}\sum_{i\alpha}c^\dagger_{i\alpha}{\bm \sigma}^{z}c_{i\alpha},
\end{eqnarray}
where $c^\dagger_{i\alpha}$($c_{i\alpha}$) is a creation (annihilation) operator for an electron with spin $\alpha$ on site $i$. The first term represents the nearest neighbor hopping term with hopping energy $t$. The second term is the effective spin-orbit coupling involving the next-nearest neighbor hopping with amplitude $t_0$. $\nu_{ij}={\vec{d_i}\times\vec{d_j}}/{|\vec{d_i}\times\vec{d_j}|}$ , where $\vec{d_i}$ and $\vec{d_j}$ are two nearest bonds connecting the next-nearest neighbor sites. The summation over $\langle...\rangle$ ($\langle\langle...\rangle\rangle$) runs over all the nearest (next-nearest) neighbor sites. The third and fourth terms are respectively the intrinsic and Rahsba spin-orbit couplings with $t_{SO}$ and $t_{R}$ the corresponding spin-orbit coupling strengths. $\hat{d_{ij}}=\vec{d_{ij}}/{|\vec{d_{ij}}|}$, where $\vec{d}_{ij}$ represents a vector from site $j$ to $i$ and $\mu_{ij}=\pm 1$ for A or B site. The last term is an exchange field $M$, which arises from the interaction with a magnetic substrate.

\begin{figure*}
  \includegraphics[width=15cm]{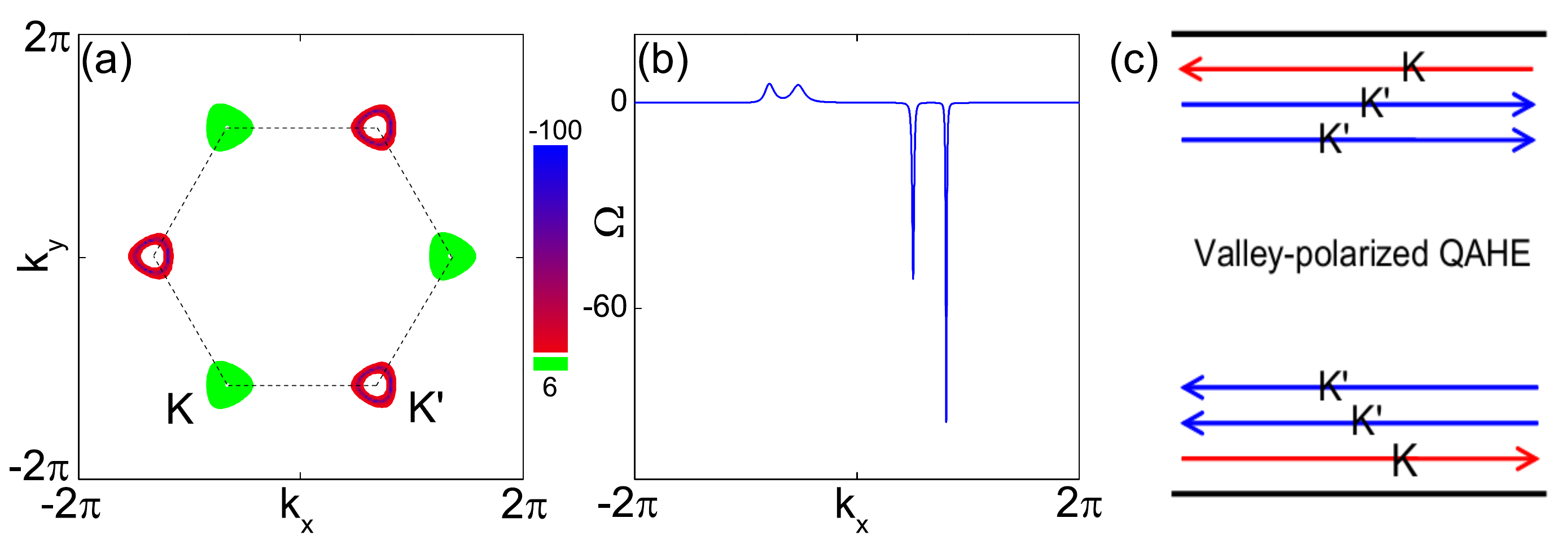}
  \caption{\label{FIG:NewPhase}(Color online) (a) Contour of Berry curvature distribution in ($k_x$, $k_y$) plane for the valley-polarized QAHE. (b) Berry curvature distribution as a function of $k_x$ at fixed $k_y=2\pi/\sqrt{3}$. (c) Valley-associated edge modes for the valley-polarized QAHE.}
\end{figure*}

By transforming the real-space Hamiltonian in Eq.~(\ref{EQ:TBH}) into a $4\times 4$ matrix $H({\bm k})$ for each crystal momentum $\bm k$ on the basis of $\{\psi_{A\uparrow},\psi_{A\downarrow},\psi_{B\uparrow},\psi_{B\downarrow}\}$, the band structure of bulk silicene can be numerically obtained by diagonalizing $H(k)$. Panel (a)-panel (f) in Fig.~\ref{FIG:Ek-bulk} plot the evolution of the bulk band structure along with the increasing of Rashba spin-orbit coupling $t_R$. Other parameters are set to be $t_0/t=0.07$, $t_{SO}/t=0.10$ and $M/t=0.50$. Thanks to the absence of intervalley scattering, valleys $K$ and $K'$ remain good quantum numbers and distinguishable. In the absence of Rashba spin-orbit coupling $t_R/t=0$, one can see that bulk band gaps open at valleys $K$ and $K'$~\cite{Ezawa1}. When the Fermi level lies inside the bulk energy gap, it is known that this insulating state is a quantum anomalous Hall insulator, which is characterized by a nonzero Chern number $\mathcal{C}$~\cite{Xiao1,Thouless,Kohmoto} calculated from
\begin{equation}
\mathcal{C}=\frac{1}{2\pi}\sum_n\int_{BZ}d^2\mathbf{k}\Omega_n,
\end{equation}
where $\Omega_n$ is the momentum-space Berry curvature for the $n$th band~\cite{Thouless,M.-C.Chang,Y.-G.Yao}
\begin{equation}
\Omega_n(\mathbf{k})=-\sum_{n'\neq n}
\frac{ 2\mathrm{Im}\langle\psi_{nk}|v_x|\psi_{n'k}\rangle
 \langle\psi_{n'k}|v_y|\psi_{nk}\rangle }{ (\varepsilon_{n'}-\varepsilon_{n})^{2} }.
\end{equation}
The summation is over all occupied valence bands in the first brilloin zone blow the bulk energy gap, and $v_{x(y)}$ is the velocity operator along x(y) direction. The absolute value of $\mathcal{C}$ corresponds to the number of gapless chiral edge states along any edge of the two-dimensional system. When $t_{R}/t=0$, the Chern number obtained from the tight-binding Hamiltonian is $\mathcal{C}=2$. Furthermore, by calculating the Chern number contribution of each valley based on a continuum model Hamiltonian, it is found that $\mathcal{C}_K=\mathcal{C}_{K'}=1$. In this case, the corresponding valley Chern number $\mathcal{C}_V=\mathcal{C}_K-\mathcal{C}_{K'}$ is vanishing. As plotted in panels (b)-(f), when the Rashba spin-orbit coupling gradually increases from $t_R/t=0.01$ to $t_R/t=0.12$, we find that the bulk band gap at valley $K$ always increases. To our surprise, the corresponding bulk band gap at valley $K'$ closes and reopens twice, signaling two possible topological phase transitions.

\begin{figure}
  \includegraphics[width=8.5cm]{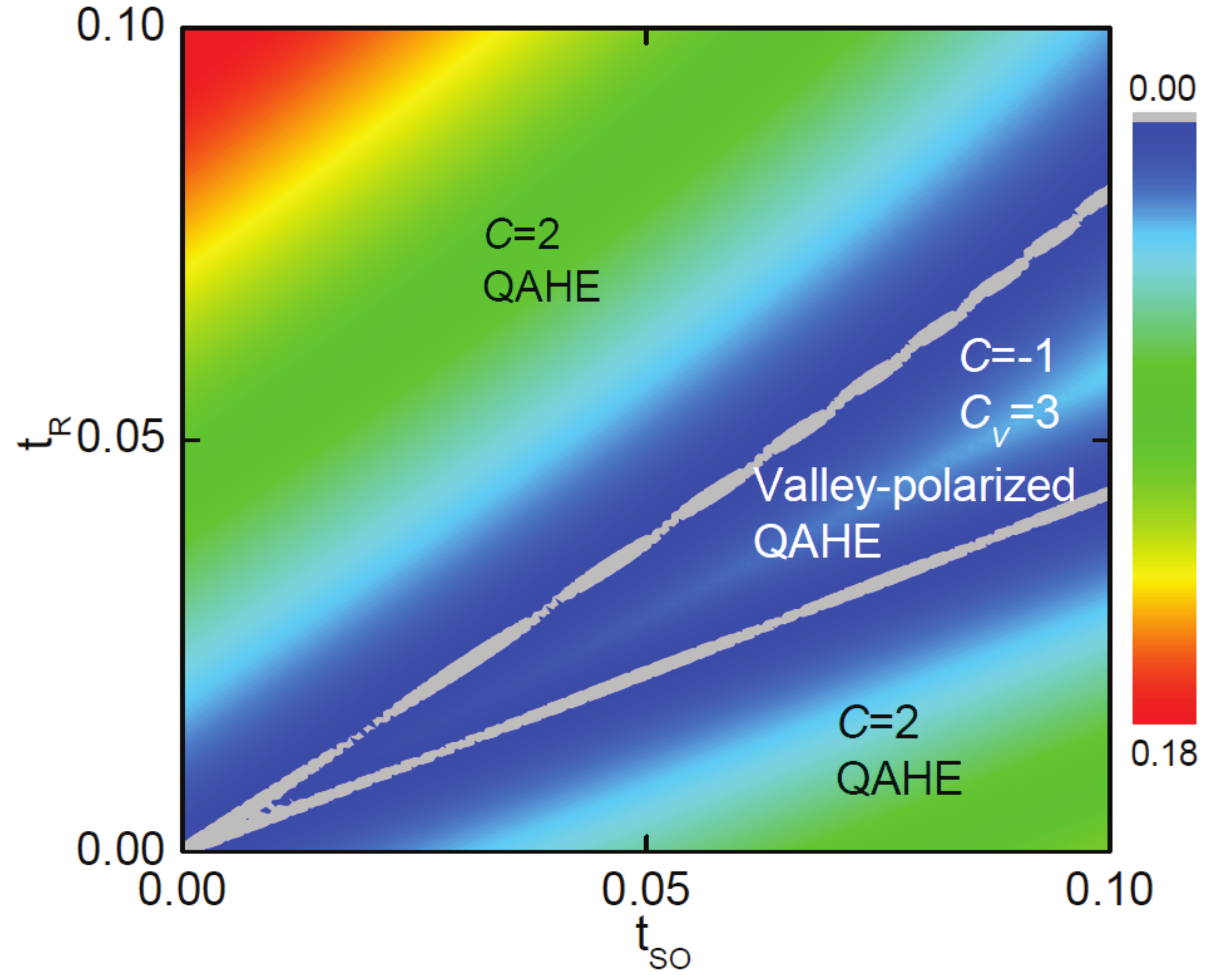}
  \caption{\label{FIG:ChernNum-tRt1}(Color online) Phase diagram in the ($t_{SO}$, $t_R$) plane. Other parameters are set to be
  $t_0/t=0.07$ and $M/t=0.50$.
  The grey lines separate three topological phases for the single layer silicene, and colors are used to indicate the size of the bulk band gap. In phase-I and phase-III, the Chern numbers are identical $\mathcal{C}=2$. While in phase-II, the Chern number is $\mathcal{C}=-1$ and $\mathcal{C}_V=3$.}
\end{figure}

Let us now explore the topological properties of the resulting two phases based on the Chern number calculation described in Eq.~(\ref{FIG:ChernNum-tRt1}). In the first phase after the gap reopening shown in Fig.~\ref{FIG:Ek-bulk}(d), the Chern number is obtained to be $\mathcal{C}=-1$. Since the bulk gap near valley $K$ does not close, the resulting Chern number keeps unchanged. However, the corresponding contribution from valley $K'$ becomes $\mathcal{C}_{K'}=-2$. Naturally, this imbalance of Chern number contribution between valleys $K$ and $K'$ gives rise to a nonzero valley Chern number $\mathcal{C}_V=3$. This means that the new topological insulating phase is both a QAHE and a QVHE. Such a state is robust against any kind of impurities (magnetic or not) and carries a net valley-polarized current. For simplicity, hereinafter we name it as a valley-polarized QAHE. To our knowledge, this is the first time such phase is reported. To further explore this new phase, in Fig.~\ref{FIG:NewPhase}(a) we show the contour of its Berry curvature distribution in the ($k_x$, $k_y$) plane. And Fig.~\ref{FIG:NewPhase} (b) plots the Berry curvature as a function of $k_x$ at fixed $k_y=2\pi/\sqrt{3}$. One can find that the Berry curvatures are mainly localized around the valley points, and obviously the Berry curvature density near valley $K$ is different from that near valley $K'$. This is a direct consequence of the inequality of the Chern number carried by valley K and K'.

In another new phase after the second topological phase transition for even larger $t_R$ [see in Fig.~\ref{FIG:Ek-bulk}(f)], the Chern number becomes $\mathcal{C}=2$ again, where $\mathcal{C}_{K'}=1$ for valley $K'$, indicating that this phase is the same as the phase without Rashba spin-orbit coupling.
In this way, we have realized a Chern-number-tunable QAHE by controlling the Rashba spin-orbit coupling strength. In addition to the Chern number, gapless edge mode inside the bulk energy gap provides a more intuitional picture to characterize the QAHE.

Figures~\ref{FIG:Ek-bulk}(g)-\ref{FIG:Ek-bulk}(l)  plot the one-dimensional band structure of zigzag-terminated silicene ribbon. One can observe that near valley $K$ there is always one pair of edge mode for any Rashba spin-orbit coupling strength. While at valley $K'$ the corresponding edge modes vary along with the topological phase transitions, i.e., after the first and second topological phase transitions there are respectively two and one pair of edge modes. Through studying the wavefunction distribution of the gapless edge states inside the bulk band gap shown in Fig.~\ref{FIG:Ek-bulk}(j) and from the energy dispersion, we find that the edge modes of the valley-polarized QAHE have the form as plotted in Fig.~\ref{FIG:NewPhase} (c): (1) there are three edge states localized at each boundary; (2) For the upper (lower) boundary, two edge states encoded with valley $K'$($K$) propagate from left (right) to right (left) while one encoded with valley $K$ ($K'$) counterpropagate from right (left) to left (right); (3) a net valley-polarized current associate with valley $K'$ ($K$) flows to the right (left) direction.

In below, we present a simple physical understanding of the formation of this valley-polarized QAHE using the spin texture picture. By analyzing the spin textures of the real-spin $\bm {s}$ and the pseudo-spin textures of AB sublattices $\bm {\sigma}$~\cite{Qiao2}, we show that the topological charges carried by the real-spin or AB sublattice pseudo-spin for valleys $K$ and $K'$ are
\begin{eqnarray}
&& n^{K}_{1\bm{s}}=n^{K'}_{1\bm{s}}=0.0; \nonumber \\
&& n^{K}_{2\bm{s}}=1.0; n^{K'}_{2\bm{s}}=-2.0; \nonumber \\
&& n^{K}_{1\bm{\sigma}}=n^{K'}_{1\bm{\sigma}}=0.5; \nonumber \\
&& n^{K}_{2\bm{\sigma}}=n^{K'}_{2\bm{\sigma}}=-0.5;
\end{eqnarray}
where `1' and `2' labels the two valence bands below the bulk band gap. From the above finding, one can conclude that the net Chern number is mainly contributed from the real-spin texture carried by the $2^{nd}$ valence band.

Finally, we provide a phase diagram in the ($t_{SO}$, $t_{R}$) plane. Figure~\ref{FIG:ChernNum-tRt1} clearly shows that there are three topological phases separated by two gray lines representing the topological phase transition. Here, we use color to signify the size of the corresponding bulk band gap. From the Chern number calculation, we show that the Chern number in phase-I is identical to that in phase-III, i.e., $\mathcal{C}=2$. However, in phase-II the Chern number is $C=-1$ and the valley Chern number is $\mathcal{C}_V=3$, which is exactly the newly found valley-polarized QAHE.

In summary, in the monolayer silicene we have found by tuning the Rashba spin-orbit coupling, a new topological phase---the valley-polarized quantum anomalous Hall state. Different from a conventional quantum anomalous Hall state, the valley-polarized quantum anomalous Hall state has a quantized Chern number $\mathcal{C}=-1$ and a nonzero valley Chern number $\mathcal{C}_V=3$. Therefore, the new topological phase possesses the properties of both quantum anomalous Hall effect and quantum valley Hall effect, which is a good candidate to design dissipationless valleytronics. The Rashba spin-orbit coupling can serve as a topological switch to drive the monolayer silicene from a conventional quantum anomalous Hall phase to a valley-polarized quantum anomalous Hall phase, which can be realized by controlling the adatom coverage in monolayer silicene.

This work was financially supported by National Basic Research Program of China
(973 Program Grants No. 2011CBA00100), the NSFC (Grant Nos. 11174022, 11174337, and 11225418),
the NCET program of MOE (Grant No. NCET-11-0774),
and the Fundamental Research Funds for the Central Universities.

\end{document}